# A COMPARATIVE STUDY OF CLUSTERHEAD SELECTION ALGORITHMS IN WIRELESS SENSOR NETWORKS

K.Ramesh[1] and Dr. K.Somasundaram[2]

[1]Dept of ECE, Nandha Engineering College, Erode.
rameshk.me@gamil.com
[2]Dept of CSE, Arupadai Veedu Institute of Technology, Chennai.
soms72@yahoo.com

## ABSTRACT

*In Wireless Sensor Network, sensor nodes life time is the most critical parameter. Many researches on these lifetime extension are motivated by LEACH scheme, which by allowing rotation of cluster head role among the sensor nodes tries to distribute the energy consumption over all nodes in the network. Selection of clusterhead for such rotation greatly affects the energy efficiency of the network. Different communication protocols and algorithms are investigated to find ways to reduce power consumption. In this paper brief survey is taken from many proposals, which suggests different clusterhead selection strategies and a global view is presented. Comparison of their costs of clusterhead selection in different rounds , transmission method and other effects like cluster formation, distribution of clusterheads and creation of clusters shows a need of a combined strategy for better results.*

## KEYWORDS

*Wireless Sensor Network, cluster-head (CH), LEACH*

## 1. INTRODUCTION

A wireless sensor network (WSN) is a wireless network consisting of spatially distributed autonomous devices using sensors to cooperatively monitor physical or environmental conditions such as temperature, sound, vibration, pressure, motion pollutants at different locations.

Wireless sensor networks consist of hundreds to thousands of low-power multi functioning sensor nodes, operating in an unattended environment with limited computational and sensing capabilities. In addition to one or more sensors, each node in a sensor network is typically equipped with a radio transceiver or other wireless communications device, a small microcontroller and an energy source, usually a battery. These inexpensive and power-efficient sensor nodes works together to form a network for monitoring the target region.

Through the co-operation of sensor nodes, the WSNs collect and send various kinds of message about the monitored environment (e.g. temperature, humidity, etc.) to the sink (base) node, which processes the information and reports it to the user.

The development of wireless sensor networks was originally motivated by military applications such as battlefield surveillance [2]. Recent developments in this technology have made these sensor nodes available in a wide range of applications in military and national security, environmental monitoring, and many other fields.

Wireless sensor networks have the following characteristics:

- ✓ It includes two kinds of nodes:
  1. *Sensor nodes* with limited energy can sense their own residual energy and have the same architecture;
  2. One *Base Station (BS)* without energy restriction is far away from the area of sensor nodes.
- ✓ All sensor nodes are immobile. They use the direct transmission or multi-hop transmission to communicate with the BS.
- ✓ Sensor nodes sense environment at a fixed rate and always have data to send to the BS.
- ✓ Sensor nodes can revise the transmission power of wireless transmitter according to the distance.
- ✓ Cluster head perform data aggregation and BS receives compressed data.
- ✓ The lifespan of WSN is the total amount of time before the first sensor node runs out of power.

In this paper Wireless Sensor Networks, sensor node and its characteristics are introduced in first section. Clustering concepts are introduced in the second section. In Section 3 brief survey results with different parameters are given and concluded in section 4.

# 2. CLUSTERING

## 2.1 Cluster Formation

Sensor nodes typically use irreplaceable power with the limited capacity, the node's capacity of computing, communicating, and storage is very limited, which requires WSN protocols need to conserve energy as the main objective of maximizing the network lifetime. An energy-efficient communication protocol LEACH, has been introduced [16] which employs a hierarchical clustering done based on information received by the BS. The BS periodically changes both the cluster membership and the cluster-head (CH) to conserve energy.

The CH collects and aggregates information from sensors in its own cluster and passes on information to the BS. By rotating the cluster-head randomly, energy consumption is expected to be uniformly distributed. However, LEACH possibly chooses too many cluster heads at a time or randomly selects the cluster heads far away from the BS without considering nodes' residual energy. As a result, some cluster heads drain their energy early thus reducing the lifespan of WSN.

In each round of the cluster formation, network needs to follow the two steps to select clusterhead and transfer the aggregated data**.** ***(i)*** Set-Up Phase, which is again subdivided in to Advertisement, Cluster Set-Up & Schedule Creation phases. (ii) Steady-State Phase, which provides data transmission using Time Division Multiple Access (TDMA).

The election of cluster head node in LEACH [16] has some deficiencies such as,

- ✓ Some very big clusters and very small clusters may exist in the network at the same time.
- ✓ Unreasonable cluster head selection while the nodes have different energy.
- ✓ Cluster member nodes deplete energy after cluster head was dead.
- ✓ The algorithm does not take into account the location of nodes.

- ✓ Ignores residual energy, geographic location and other information, which may easily lead to cluster head node will rapidly fail.

Motivated from this, so many clustering proposals are reported in the literature, suggesting different strategies of clusterhead selection and its role rotation.

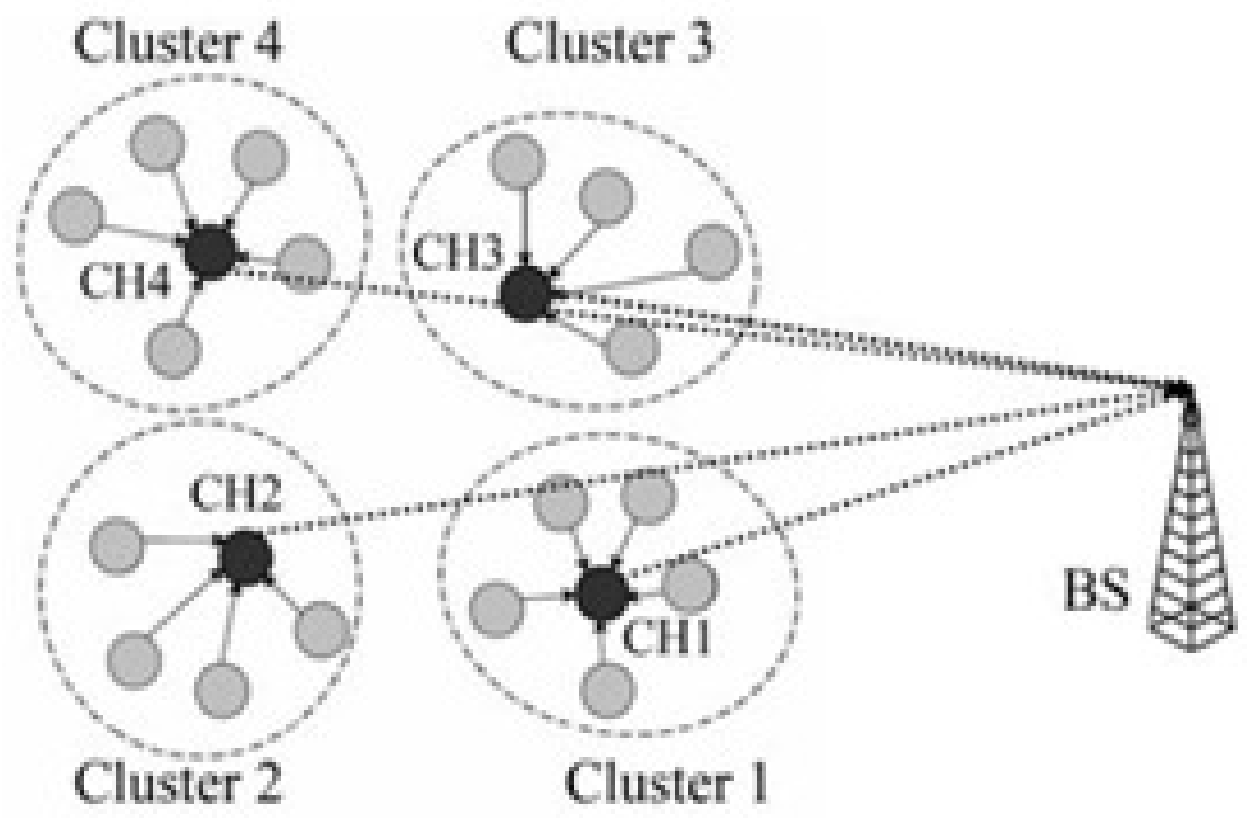

**Fig. 1.** The LEACH clustering communication hierarchy for WSNs.

To have a global view of these strategies of clusterhead selection, their necessary characterization on a common platform raises following questions.

- ✓ Who initiates the clusterhead selection?
- ✓ Which parameters decide the role of a sensor node?
- ✓ Which sensor nodes shall be selected as clusterheads?
- ✓ Does it require re-initiation of cluster formation process?
- ✓ Are the selected clusterheads evenly distributed?
- ✓ Does it guarantee creation of balanced clusters?
- ✓ Which method is suitable for large network, Single-hop or Multi-hop?

## 2.2 WSNs Topologies

WSN topologies are classified into four types of models as shown in Fig. 2. In the single-hop models (Fig. 2(a) and Fig. 2(b)), all sensor nodes transmit their data to the sink node directly. These architectures are infeasible in large-scale areas because transmission cost becomes expensive in terms of energy consumption and in the worst case, the sink node may be unreachable.

In the multi-hop models, we can consider the flat model (Fig. 2 (c)) and the clustering model (Fig. 2(d)). In the multi-hop flat model, because all nodes should share the same information such as routing tables, overhead and energy consumption can be increased. On the other hand, in the multi-hop clustering model, sensor nodes can maintain low overhead and energy consumption because particular cluster heads aggregate data and transmit them to the sink node. Additionally, wireless medium is shared and managed by individual nodes in the multi-hop flat model, which results in low efficiency in the resource usage. In the multi-hop clustering model, resources can be allocated orthogonally to each cluster to reduce collisions between clusters and be reused cluster by cluster. As a result, the multi-hop clustering model is appropriate for the sensor network deployed in remote large-scale areas.

### 2.3 Clustering Strategies – Classification

While improving the limitations of LEACH, many clustering proposals for increasing network lifetime are reported suggesting different strategies of clusterhead selection and its role rotation among the sensor nodes, using different parameters. Based on these parameters, these strategies of clusterhead selection may broadly be categorized as *deterministic, adaptive* and *combined metric (hybrid).*

In *deterministic schemes* special attributes of the sensor node such as their identification number (Node ID), number of neighbours they have (Node degree) and in *adaptive schemes* the resource information like remnant energy, energy dissipated during last round , initial energy of the nodes are used to decide their role during different data gathering rounds.

Based on who initiates the clusterhead selection, the adaptive schemes may be categorized as base station assisted or self organized (Probabilistic). Based on the parameters considered for deciding the role of a sensor node, the probabilistic schemes may further be classified as fixed parameter or resource adaptive.

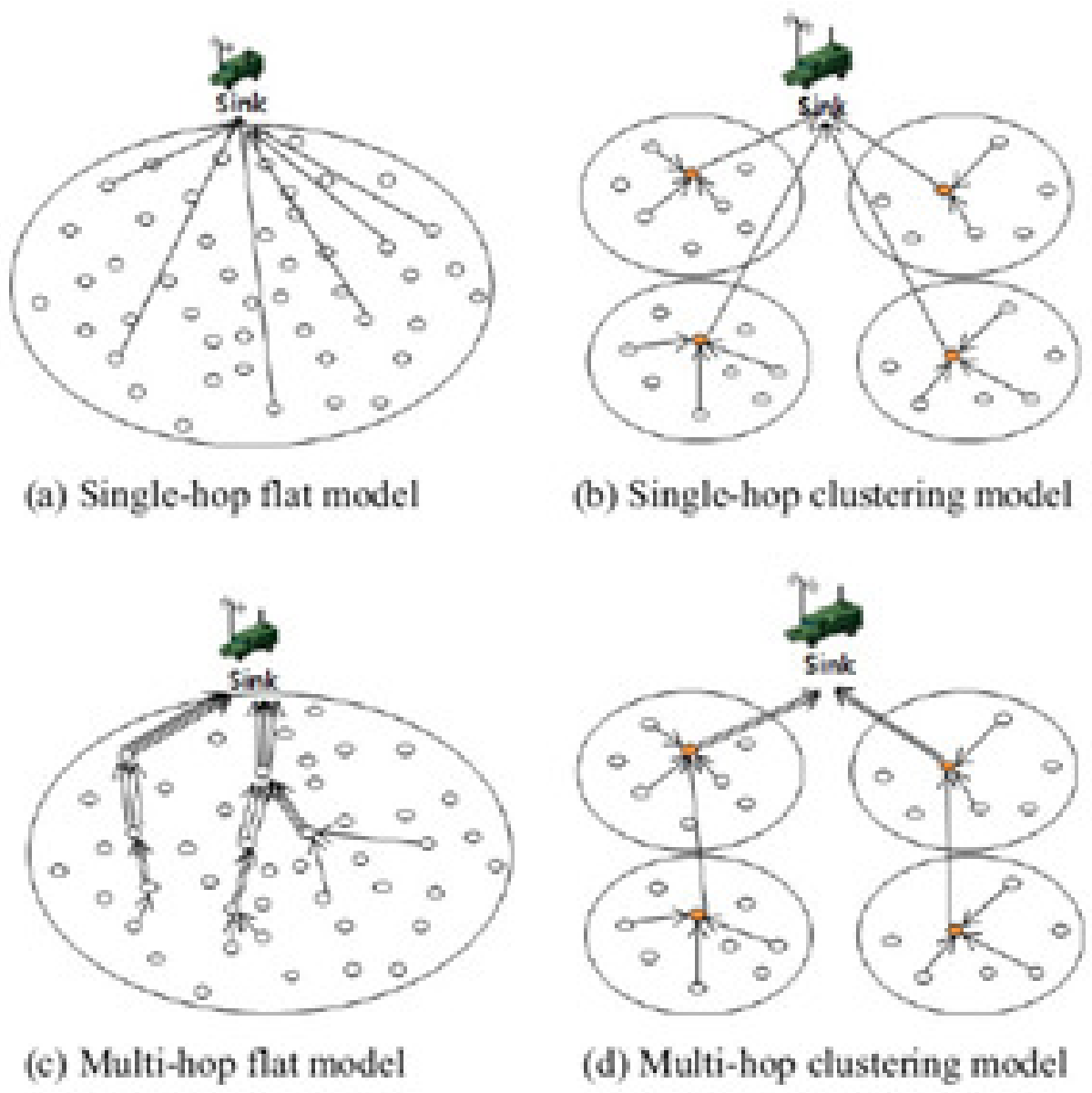

Fig 2. Classification of WSN Topologies

Few proposals, reported, use combination of deterministic and adaptive approaches and may be termed as combined metric (hybrid) schemes.

## 3. COMPARISON OF VARIOUS CLUSTERHEAD SELECTION STRATEGIES

Comparison of various clusterhead selection strategies in terms of their assistance considered in clusterhead selection (CSA), parameters used, required re-clustering (RC), required cluster formation (FC), even or fair distribution of clusterheads (DCH), and creation of balanced clusters (BCC) seems to be meaningful, to have their broader understanding.

## 3.1 Deterministic Schemes

In a communication range, sensor nodes first satisfying the fixed node degree criterion select themselves as clusterheads. To decide on the clusterhead role, during each round, all sensor nodes broadcast hello message to their neighbours and the nodes first receiving as many as pre-defined number of these broadcasts declare themselves as clusterheads and broadcast a cluster setup. Existence of exactly one clusterhead is ensured in a communication range by not allowing the sensor nodes receiving the setup broadcast to broadcast again. The sensor nodes receiving the setup broadcast then send the joining requests and the clusterhead after receiving these requests confirms the joining, prepares and distributes the time schedule for its cluster members.

ACE-C [7], for even distribution of sensor nodes and to avoid re-clustering during each round, select clusterheads for each round based on node ID's. Initially all sensor nodes are assigned ID's from 0 to N-l (N is the number of nodes in the network). Depending on the number of clusterheads (C) required for each round the necessary number of nodes (with ID's from 0 to C-l) are selected as clusterheads for first round. For next round the nodes with ID's from C to 2C-1 are selected as clusterheads. To distribute the clusterheads evenly over the network, ACE-L [4] uses location information, provided in the form of reference points, to decide the clusterhead during each round.

Based on number of clusterheads required equal number of reference points is fixed, a priori. The nearest among these points is used as a main reference point (MRP) by the sensor nodes. Nodes with same MRP values contend for the role and the one with minimum delay elects itself as a clusterhead for current round The nodes receiving the selection beacon from this clusterhead leave the competition and join it as cluster members. However both ACE-C and ACE-L needs clusters to be formed after each role rotation of clusterhead.

Deterministic clusterhead selection strategies discussed in this section are compared, below in Table 1 with respect to their requirements of clusterhead selection and the associated after effects.

Table 1: Comparison of Deterministic Schemes

| **Scheme** | **CSA** | **Parameters** | RC | **DCH** | **BCC** |
|---|---|---|---|---|---|
| ACE-C [4] | Sensor nodes (Self organized) | Node ID | NO | NO | NO |
| ACE-L [4] | | Reference Point (MRP) | YES | YES | YES |
| RCLB [5] | | Number of CHs range | YES | YES | NO |

## 3.2 Base Station Assisted Adaptive Schemes

The base station, based on the node deployment information either priori available or collected from the sensor nodes, clusters the network and informs it to these nodes. The clusterheads are either elected by the base station or selected by the sensor nodes.

Particularly in LMSSC [6], the network is first partitioned into clusters by the base station and appropriate number of clusterheads are decided by evaluating a node metric which is defined, for any sensor node, as a ratio of its residual energy to the aggregate of sum of squared distances from a concerned sensor node to every other sensor node in the cluster and its squared distance to the base station.

All sensor nodes communicate their position information and energy level to the base station in LEACH-C [7] and provide the necessary information to calculate the average node energy. Sensor nodes with remaining energy below this value are restricted from becoming clusterhead during current round. Base station finds the predefined number of clusterheads and divides the network into clusters, so as to minimize the energy required for non clusterhead members to transmit their data to the clusterhead. However formation of clusters with equal number of nodes in each of them is not guaranteed with this scheme.

To avoid re-clustering, LEACH-F [9] uses a stable cluster and rotating clusterhead concept in which cluster once formed is maintained stable, throughout the network lifetime. Only the responsibility of cluster data gathering is rotated within the nodes in the cluster. Initially the clusterheads are selected and clusters are formed using LEACH-C algorithm.

Other base station assisted schemes are Controlled Density Aware Clustering Protocol (CBCDACP) [10] where the base station centrally performs the cluster formation task, Two-Tier Data Dissemination approach [11] that provides scalable and efficient data delivery with location aware, FZ-LEACH [12] that forming Far-Zone which is a group of sensor nodes which are placed at locations where their energies are less than a threshold, In Adaptive Cluster Head Election and Two-hop LEACH protocol (ACHTHLEACH) [13], Nodes are tagged as near nodes or far nodes according to the distances to the BS. The near nodes belong to one cluster while the far nodes are divided into different clusters by the Greedy K-means algorithm. The cluster head is shifted and the node with the maximal residual energy in each cluster is elected.

In document [14] a cluster head election called Grid Sectoring base-on distribution of load balancing and energy consumption over both uniform and non-uniform deployment were presented and in Optimal Placement of Cluster-heads (OPC) algorithm [15], the key future is handling the load near the sink is to vary the density and the transmission range of the cluster-heads based on the distance between cluster-heads and the sink.

Base station assisted schemes are compared with respect to different features of their clusterhead selection, below in Table 2. It should be noted that the base station is responsible for re-clustering in most of these schemes and the sensor nodes do not perform the computations for selecting the clusterheads.

### 3.3 **Fixed parameter probabilistic schemes**

In these schemes, clusterheads are selected for initial and subsequent data gathering rounds by evaluating an expression involving some probabilistic requirements, utilizing fixed parameters like number of clusterheads and round number.

In LEACH, clusterhead role is rotated among all sensor nodes by re-clustering the network after specific number of data gathering cycles called *round.* During each round, a fixed percentage of total network nodes are selected as clusterheads which then start cluster formation process by advertising their selection to the non clusterhead nodes that on receipt of these equal transmit power advertises, from different clusterheads, join one with highest received signal strength. Each node in the network chooses a random number between 0 and 1 and if this number is less than the evaluated adaptive threshold, selects itself as clusterhead for the current round.

In LEACH, during some round, it is possible that none of the node selects itself as clusterhead and all the nodes have to act as forced clusterheads. To improve upon, such a round is treated as cancelled in power efficient communication protocol [17] and a fresh clusterhead selection is carried out, independent of the current round.

**Table 2.** Comparison of base station assisted schemes

| Scheme | CSA | Parameters | RC | DCH | BCC |
|---|---|---|---|---|---|
| LMSSC[6] | Base station | Residual energy, CH to SN and CH to BS distances | NO | YES | YES |
| LEACH-C [7] | | Position information and energy level " | NO | NO | NO |
| BCDCP [8] | | Position information and energy level | NO | YES | YES |
| LEACH-F [9] | | Position information | NO | NO | YES |
| CBCDACP) [10] | | Min distance b/n node to base | NO | YES | YES |
| TTDD[11] FZ-LEACH[12] | | Location aware | NO | YES | YES |
| [13], [14], [15] | | Optimal placement | NO | NO | YES |

ERA [18] suggests an improvement in cluster formation phase in which the non clusterhead nodes while deciding the clusterhead to join select a path with maximum sum of residues. The strategy proposed in [19] uses CDMA codes from cluster set up whereas LEACH uses them during data gathering. The TDMA schedule is distributed immediately after confirming joining, on receipt of joining request from a node, to avoid collision during node's reply to clusterhead advertise, to achieve the energy efficiency. RRCH in [22] in which, initial clusterhead selection and cluster formation is carried out following LEACH algorithm and TB-LEACH [23] suggests to select constant number of clusterheads, the partition of clusterhead is balanced and uniform.

Following Table 3 summarizes, various features associated with, the clusterhead selection and role rotation strategies discussed in this section.

**Table 3:** Comparison of fixed parameter probabilistic schemes

| Scheme | CSA | Parameters | RC | CF | DCH | BCC |
|---|---|---|---|---|---|---|
| LEACH [16], [17],[18], [19], [20], [21], [23] | Sensor nodes | Number of clusterheads, round number | YES | YES | NO | NO |
| RRCH[22] | | | NO | NO | NO | NO |

## 3.4 Resource adaptive probabilistic schemes

In resource adaptive schemes, information about the available node resources is utilized, while selecting clusterheads for the subsequent rounds.

The scheme suggested in [24] calculates the threshold considering residual energy, energy dissipated during current round and average node energy as additional parameters and makes the clusterhead selection strategy energy adaptive. The nodes in the network take decision about their clusterhead role carrying out a process similar to LEACH but with a resource adaptive threshold value. LEACH-B [25], Energy-LEACH [26] and scheme in [27] also adapt the LEACH threshold using different energy values.

In HEED [28], sensor nodes use residual energy as a criterion to decide on their role as a clusterhead and make up their mind setting the probability to a value expressed in terms of residual energy, maximum energy and the optimum percentage of clusterheads required for a particular data gathering round which is not allowed to fall below a minimum pre-defined threshold. In schemes [30], power optimized LEACH [31], ALEACH [32], EAMC [33], EAP [34], CEFCHS [35], FRCA [36], LEACH-M [37] are suggests that node's remaining energy or residual energy as the main constrain to select a node as clusterhead.

The clusterhead selection strategies discussed in this section are summarized in Table 4, below.

**Table 4:** Comparison of resource adaptive probabilistic schemes

| Scheme | CSA | Parameters | RC | CF | DCH | BCC |
|---|---|---|---|---|---|---|
| [24] | Sensor nodes (Self organized) | Energies – residual, in last round. | YES | YES | NO | NO |
| LEACH-B [ 25] | | Energy | YES | YES | NO | NO |
| Energy-LEACH [26] | | Energy | YES | YES | NO | NO |
| [27] | | Remnant Energy | YES | YES | NO | NO |
| HEED [28] | | Energies-residual and optimum percentage of CH | YES | YES | NO | NO |
| LEACH-ET [29] | | Threshold | YES | YES | NO | NO |
| [30] to [37] | | Residual energy in the range | YES | YES | YES | YES |

## 3.5 Clusterhead Selection in Hybrid Clustering (Combined Metric) Schemes

In cluster based data gathering literature, some hybrid approaches are suggested combining clustering with, one or more of the, other architectures and increased energy efficiency is claimed. In M-LEACH [38] that adjusting the nodes, Threshold function, when non cluster-heads choose optimal cluster-head, they consider comprehensive nodes' residual energy and distance to base-station, then compare their performance, the simulation results show that the new strategy of cluster-heads election achieve great advance in sensor and in ACAER [39] which periodically selects cluster nodes according to their coverage rate and residual energy.

The EAMC [40] can reduce the number of relays used for data transmission by minimizing the amount of the nodes in the root tree (that is cluster-head). Unequal Cluster-based Routing (UCR) [41] protocol groups the nodes into clusters of unequal sizes. Cluster heads closer to the base station have smaller cluster sizes than those farther from the base station, thus they can preserve some energy for the inter-cluster data forwarding and [42] using decision tree algorithm to select the best node as a cluster head.

Gradual Cluster head election Algorithm (GCA) [43] which gradually elects cluster heads according to the proximity to neighbour nodes and the residual energy level and one-hop neighbour information (GCA-ON), which elects cluster heads based on *Er* and the relative location information of sensor nodes. LEACH-improve [44] consider both energy and coverage together.

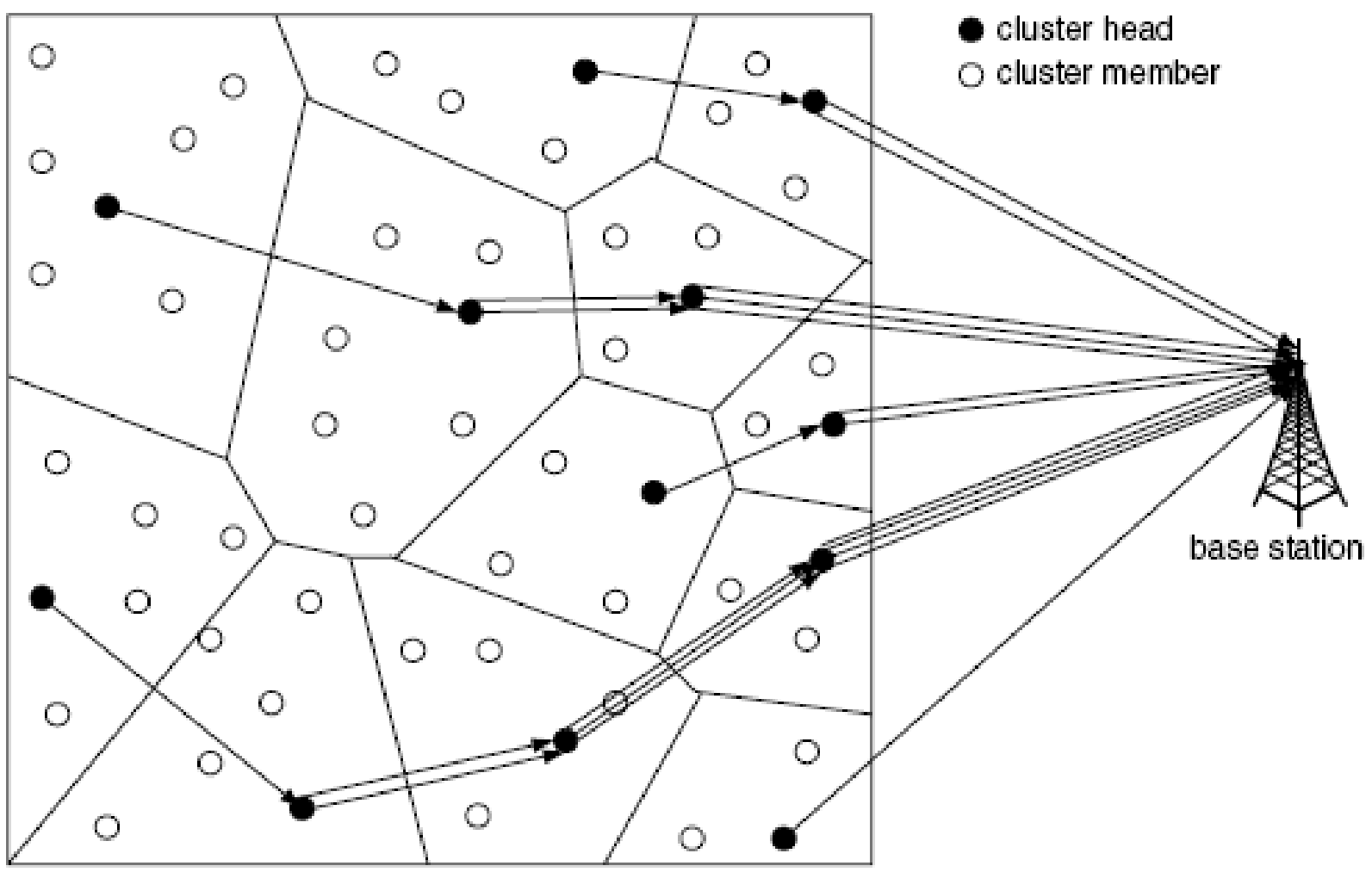

**Fig. 3. An overview of the UCR protocol**

Particularly, Concentric Clustering Scheme proposed in PEGASIS [45] extend chain formation approach to clustering architecture for data gathering in wireless sensor networks. In concentric clustering scheme, each node in the sensor network assigns itself a layer number, based on the received signal strength (RSSS) of the control message from the base station. The nodes with same layer numbers form a chain in their respective layers and one of these nodes is selected as a head node (clusterhead) for that layer. The role of head node is rotated among the nodes in the same layer.

**Table 5:** Comparison of Hybrid Clustering Schemes

| Scheme | CSA | Parameters | RC | CF | DCH | BCC |
|---|---|---|---|---|---|---|
| M-LEACH[24] | Sensor nodes (Self organized) | Threshold, location | NO | YES | YES | YES |
| ACAER [39], EAMC [40], UCR [41]*, [42], GCA [43], LEACH-imp [44] | | Coverage rate and residual energy | NO | YES | NO | YES |
| PEGASIS [45] | Base station | RSSS | NO | NO | YES | YES |

***- Additional parameter unequal cluster size***

# 4. CONCLUSION

In this paper, different clustering schemes are classified and discussed with special emphasis on their clusterhead selection strategies. They are compared with respect to their requirement of (1) clustering during each round for selecting the clusterheads, (2) cluster formation required after each rotation of role of clusterhead, (3) distribution of clusterheads over the network, (4) creation of balanced clusters, (5) parameters used and (6) the assistance considered to highlight the effect of clusterhead selection strategy on the performance of these schemes. This survey also answers all the questions raised at starting stage about the clustering and clusterhead selection sections. The use of these parameters for this comparison is justified by reasoning the effects of clusterhead selection and its role rotation on the energy efficiency of the network. For multi-hop data forwarding, from clusterhead to base station, distance between the forwarding clusterhead and intermediate clusterhead shall be maintained approximately same, during different data gathering rounds, to ensure equal amount of energy consumption due to their data forwarding to or towards the base station. Finally it is concluded from the survey that, still it is needed to find more scalable, energy efficient and stable clustering scheme, for data gathering in wireless sensor networks.

# REFERENCES

[1] O. Younis, M. Krunz and A. Ramasubramanian, "Node Clustering in Wireless Sensor Networks: Recent Developments and Deployment Challenges", *IEEE Network,* May-June-2006, pp. 20-25.

[2] 1. Yu and P. Chong, "A survey of clustering schemes for Mobile Ad Hoc Networks", *IEEE Communications Surveys and Tutorials,* First Quarter,
2005, Vol 7, No.1, pp. 32-47.

[3] A. Abbasi and M. Younis, "A survey on clustering algorithms for wireless sensor networks", *Elsevier Sci, J. Computer Communications,* Vo1.30,
2007, pp. 2826-2841.

[4] C. Liu, C. Lee, L.ChunWang, "Distributed clustering algorithms for data gathering in wireless mobile sensor networks", *Elsevier Sci. J. Parallel Distrib. Comput.,* Vol.67, 2007, pp.l187 -1200

[5] N.Kim, J.Heo, H.Kim and W. Kwon, "Reconfiguration of clusterheads for load balancing in wireless sensor networks" *Elsevier Sci. J. Compo Communications,* Vol. 31,2008, pp. 153-159

[6] P. Tillaport, S. Thammarojsakul, T.Thumthawatworn and P. Santiprabhob, "An Approach to Hybrid Clustering and Routing in Wireless Sensor Networks", In Proc. *IEEE Aerospace,* 2005, pp. 1-8

[7] W. Heinzelman, A. Chandrakasan and H. Balakrishnan., "An Application-Specific Protocol Architecture for Wireless Microsensor Networks", *IEEE Trans. Wireless Communications,* Vol. 1, No.4, October 2002, pp.660-670.

[8] S.Muruganathan, D. Ma, R.Bhasin and A. Fapojuwo, "A Centralized Energy-Efficient Routing Protocol for Wireless Sensor Networks", *IEEE Radio Communication,* March, 2005, pp. S8-Sl3

[9]W. Heinzelman, *Application-Specific Protocol Architectures for Wireless Networks,* Ph.D Thesis, Massachusetts Institute of Technology, June 2000.

[10] Jannatul Ferdous, Mst. Jannatul Ferdous, and Tanay Dey, "A Comprehensive Analysis of CBCDACP in Wireless Sensor Networks", Journal of Communications, VOL. 5, NO. 8, August 2010.

[11]Fan Ye, Haiyun Luo, Jerry Cheng, Songwu Lu, Lixia Zhang "A Two-Tier Data Dissemination Model for Large scale Wireless Sensor Networks"ACM, MOBICOM'02, September 23–28, 2002.

[12]Vivek Katiyar, Narottam Chand "Improvement in LEACH Protocol for Large-scale Wireless Sensor Networks" Proceedings of ICETECT 2011

[13]Li-Qing Guo, Yi Xie, Chen-Hui Yang, Zheng-Wei Jing "Improvement on LEACH by Combining Adaptive Cluster Head Election and Two-Hop Transmission" Proceedings of the Ninth International Conference on Machine Learning and Cybernetics, Qingdao, 11-14 July 2010

[14]Anirooth Thonklin, W. Suntiamorntut "Load Balanced and Energy Efficient Cluster Head Election in Wireless Sensor Networks" 8$^{th}$ ECTI Conference 2011

[15]M. Dhanaraj and C. Siva Ram Murthy "On Achieving Maximum Network Lifetime through Optimal Placement of Cluster-heads in Wireless Sensor Networks" ICC 2007 proceedings.

[16] W..Heinzelman, A. Chandrakasan and H. Balakrishnan, "Energy-Efficient Communication Protocol for Wireless Microsensor Networks", In Proc. *33rd HICS,* 4-7 Jan, 2000, Vol 2, pp. 10.

[17] C. Liu and C. Lee, "Power Efficient Communication Protocols for Data Gathering on Mobile Sensor Networks", In Proc. *IEEE Int. Con!Vehicular Technology (VTC'04),* 2004, pp. 4635-4639.

[18] H.Chen, C. Wu, Y. Chu, C. Cheng and L. Tsai, "Energy Residue Aware (ERA) Clustering Algorithm for Leach-based Wireless Sensor Networks", In Proc. *2nd Int. Con! Systems and Networks Communications (ICSNC 2007),25-31* Aug, 2007, pp.40.

[19] 1. Zhao, A. Erdogan and T. Arslam, "A Novel Application Specific Network Protocol for Wireless Sensor Networks",. In Proc. *ISCAS (6), 2005,pp.5894-5897*

[20] H. Lim, Sung Kim, H. Yeo, Seung Kim, and K. Ahn "Maximum energy routing protocol based on strong head in Wireless Sensor Networks", In Proc. *Int. Con! ALPIT.,* 2006, pp. 414-419

[21] M. Halgamuge, K Ramamohanrao and M. Zukerman," High Powered Cluster heads for Extending Sensor Network Lifetime", In *Proc. of the IEEE Int. Symposium Signal Processing and Information Technology,* Aug, 2006, pp. 64-69..

[22] D. Nam and H. Min, "An Efficient Ad-Hoc Routing Using a Hybrid Clustering Method in a Wireless Sensor Network" In Proc. *3rd IEEE WiMob '07,8-10* Oct, 2007, pp. 60.

[23]Hu Junping, Jin Yuhui, Dou Liang "A Time-based Cluster-Head Selection Algorithm for LEACH", ISCC 2008. IEEE Symposium on Computers and Communications, 2008.

[24] L.Ying and Y. Haibin, "Energy Adaptive Clusterhead Selection for Wireless Sensor Networks", In Proc. 6th *Int. Con! Parallel and Distributed Computing, Applications and Technologies (pDCAT), 5-8* Dec., 2005, pp. 634-638.

[25] A. Depedri, A. Zanella and R. Verdone, "An Energy Efficient Protocol for Wireless Sensor Networks" *In Proc. AINS,* June, 2003, pp. 1-6.

[26] F. Xiangning and S. Yulin, "Improvement on LEACH Protocol of Wireless Sensor Network", In Proc. *2007 Int. Con! Sensor Technologies and Applications,* 2007, pp. 260-264.

[27] F. Yiming and YJianjun , "The Communication Protocol for Wireless Sensor Network about LEACH", In Proc. *International Conference of Computational Intelligence and Security Workshops,* 2007, PP- 550-553.

[28] O.Younis and S. FahmY,"HEED: A Hybrid, Energy-Efficient, Distributed Clustering Approach for Ad Hoc Sensor Networks", *IEEE Trans. Mobile Computing,* Volume 3, Issue 4, Oct., 2004, pp.366-379.

[29] L. Lijun, W. Hongtao and C.Peng, "Discuss in Round Rotation Policy of Hierarchical Route in Wireless Sensor Networks", In Proc. *IEEE Int. Conf WiCOM'06, 2006,* pp. 1-5.

[30]Hamid Daneshvar Tarigh, Masood Sabaei "A New Clustering Method to Prolong the Lifetime of WSN", 3rd International Conference on Computer Research and Development (ICCRD), 2011 -Volume: 1.

[31]Chandan Maity, Chaitanya Garg, Sourish Behera "Adaptive Cluster Head characterization in LEACH protocol for power optimization in WSN" Proceedings of ASCNT-2011.

[32]Md. Solaiman Ali, Tanay Dey, and Rahul Biswas "ALEACH: Advanced LEACH Routing Protocol for Wireless Micro-sensor Networks" International Conference on Electrical and Computer Engineering, 2008. ICECE 2008.

[33]Xinfang Yan, Jiangtao Xi , Joe F. Chicharo and Yanguang Yu, "An Energy-Aware Multilevel Clustering Algorithm for Wireless Sensor Networks", International Conference on Intelligent Sensors, Sensor Networks and Information Processing, 2008. ISSNIP 2008.

[34]Ming Liu, Jiannong Cao, Guihai Chen and Xiaomin Wang "An Energy-Aware Routing Protocol in Wireless Sensor Networks", Sensors 2009 at www.mdpi.com/journal/sensors.

[35]Desalegn Getachew Melese, Huagang Xiong, Qiang Gao "Consumed Energy as a Factor For Cluster Head Selection in Wireless Sensor Networks" 6th International Conference on Wireless Communications Networking and Mobile Computing (WiCOM), 2010.

[36]Chongdeuk Lee and Taegwon Jeong "FRCA: A Fuzzy Relevance-Based Cluster Head Selection Algorithm for Wireless Mobile Ad-Hoc Sensor Networks" Sensors 2011, at www.mdpi.com/journal/sensors.

[37]Xu Long-long and Zhang Jian-jun "Improved LEACH Cluster Head Multi-hops Algorithm in Wireless Sensor Networks" Ninth International Symposium on Distributed Computing and Applications to Business, Engineering and Science – 2010.

[38]Yuhua Liu, Yongfeng Zhao, "A New Clustering Mechanism Based On LEACH Protocol", International Joint Conference on Artificial Intelligence, 2009.

[39]Xuehai Hu, DaiRong Ren, Houjun Wang, Yougang Qiu, Cheng Jiang "Adaptive Clustering Algorithm Based on Energy Restriction", Fourth International Conference on Intelligent Computation Technology and Automation, 2011.

[40]Sandip Kumar Chaurasiya, Tumpa Pal "An Enhanced Energy-Efficient Protocol with Static Clustering for WSN", International Conference on Information Networking (ICOIN), 2011.

[41]Guihai Chen · Chengfa Li · Mao Ye · JieWu "An unequal cluster-based routing protocol in wireless sensor networks" Springer Science + Business Media, LLC 2007

[42]Ghufran Ahmed and Noor M Khan "Cluster Head Selection Using Decision Trees for Wireless Sensor Networks" International Conference on Intelligent Sensors, Sensor Networks and Information Processing, 2008.

[43]Sang Hyuk Lee, Soobin Lee, Heecheol Song, and Hwang Soo Lee "Gradual Cluster Head Election for High Network Connectivity in Large-Scale Sensor Networks" Feb. 13~16, 2011 ICACT2011

[44]Yaqiong Wang*, Qi Wang, Ziyu Jin, Navrati Saxena "Improved Cluster Heads Selection Method in Wireless Sensor Networks" IEEE/ACM International Conference on Green Computing and Communications – 2010.

[45]Lindsey, S.; Raghavendra, C. PEGASIS: Power-Efficient gathering in sensor information systems. In *Proceeding of IEEE Aerospace Conference*, 2002; volume 3, 1125-1130.